\documentclass[twocolumn,showpacs,preprintnumbers,amsmath,amssymb]{revtex4}
\usepackage{graphicx}% Include figure files
\usepackage{dcolumn}% Align table columns on decimal point
\usepackage{bm}% bold math

\begin{document}
\preprint{}
%\preprint{APS/123-QED}

\title{Sensitivity of the parameters measured in pp collisions \\
 on the gluon PDF.}
\author{E. Cuautle}
\email{ecuautle@nucleares.unam.mx}
\author{A. Ortiz}
 \email{antonio.ortiz@nucleares.unam.mx}
\author{G. Pai\'c}
 \email{guypaic@nucleares.unam.mx}
\affiliation{Instituto de Ciencias Nucleares, Universidad Nacional Aut\'onoma de M\'exico \\
Apartado Postal 70-543, M\'exico Distrito Federal 04510, M\'exico.}%
%\protect\\

\date{\today}% It is always \today, today,

\begin{abstract}
The  sphericity distribution in  minimum bias  events obtained  with 
PYTHIA  event   generator,  incorporating  different  parton
distribution functions is presented. The results show that for minimum
bias pp  collisions the  sphericity distribution for  different parameters
like  the mean  transverse  momentum and  multiplicity exhibit  strong
sensitivity  on the  parton  distribution function  used. The  results
indicate that  early data  at the LHC  could be  used in this  type of
analysis to fix the gluon distribution function in the proton.

\end{abstract}

\pacs{24.85.+p, 12.38.-t}

\maketitle

\section{\label{sec:level1}Introduction }
 The calculation of the  production cross sections at hadron colliders
 relies upon a knowledge of  the distribution of the momentum fraction
 $x$ of  the partons (quarks and  gluons) in a proton  in the relevant
 kinematic range, and the corresponding fragmentation functions.  Most
 probably, at LHC, we will  be confronted with a copious production of
 jets what  will severely limit  the exclusive analysis  of individual
 jets.  It  seems therefore interesting to separate the events
 in function of their {\it multi-jetiness} and to study the parameters
 of the collisions \emph{e.g.}  mean $p_{t}$, mean multiplicity, etc.,
 in  function  of  different  parton  distribution  functions  (PDF's)
 present  on  the   market,  which as quoted by the  CTEQ  group   allows  for  an
 uncertainty   of   about   10-20\%~\cite{cteq0}.  Obviously   this
 uncertainty  will  reflects   in  the  quantification  of  the
 kinematical  processes.   The  PDF's  are of  importance  to  analyze
 standard model  processes at LHC  energies~\cite{hung-liang}, and the
 usual  leading order event  generators use  the PDF's  as an  input to
 evaluate hard  subprocess matrix elements.  In  those calculations the
 multiple parton interactions that make  up the bulk of the underlying
 event, require extensive tuning which depends strongly on the details
 of the  input parton distributions.\\ The Event  Shape Analysis (ESA)
 has   been  applied   successfully  to   the   restricted  acceptance
 experiments  at RHIC  and ALICE  to survey  the possibilities  of the
 method    to   analyze    and   isolate    specific    jet   topology
 events~\cite{esa:200a}. In the present  paper we are investigating the
 bulk properties of the event shape variables obtained in minimum bias
 pp events: the sensitivity  obtained at different collisions energies
 focussing to minimum bias collisions, corresponding to the
 minimum  \emph{x-Bjorken} ($x_B$)  achievable. The  article  is organized  as
 follows:  in the section~\ref{sec:II}  we give  the relation  for the
 sphericity  and recoil  distributions; in  the  section~\ref{PDFs} we
 discuss  the  parton distribution  functions  used  and  the way  the
 simulations were  done in the  framework of minimum bias  PYTHIA. The
 results  are  presented  in  section~\ref{Results}  and  finally,  in
 section~\ref{conclusion} the conclusion are drawn.

\section{\label{sec:II}Event shape variables}
Event shape  variables allow us  to measure geometrical  properties of
QCD final state~\cite{banfi}. Those variables has been used to analyze
dijet  production  in  hadron-hadron   collisions  and  to  study  the
sensitivity of physical parameters~\cite{banfi}.
The  shape variables at  hadron colliders  are defined  over particles
within the acceptance of the detector.  The thrust ($T$) is defined in
the transverse plane as follows:
\begin{equation}
T\equiv \underbrace{max}_{\overrightarrow{n}_{t}}
\frac{\sum_{i}|\overrightarrow{p}_{t,i}\cdot\overrightarrow{n}_{t}|}{\sum_{i}|\overrightarrow{p}_{t,i}|}
\end{equation}
where the  sum runs over all  particles in the final  state within the
acceptance,   $\overrightarrow{p}_{t,i}$   represent   the   transverse
component    of    the    momenta    of    emitted    particles    and
$\overrightarrow{n}_{t}$ is  the transverse vector  that maximizes the
ratio.  In the literature it is more common to find the quantity $1-T$
related to the transverse sphericity of the event. In the present work
we do  reference to  this as sphericity.   The range of  sphericity is
between  0 (for events  with narrow  back-to-back jets)  and $1-2/\pi$
(for events with a uniform momentum distribution).

\noindent
Another  interesting observable   is  the recoil, defined as:
\begin{equation}
R\equiv \frac{1}{\sum_{i}|\overrightarrow{p}_{t,
i}|}|\sum_{i}\overrightarrow{p}_{t i}|.
\label{ecrecoil}
\end{equation}
It is  an  indirect  measurement for  the  radiation which  remains
undetected for  acceptance effects (high inbalance: $R  \to 1$, small
inbalance: $R  \to 0$). 

\section{\label{PDFs}The parton distribution functions}
The  QCD factorization  theorem  provides us  with  a framework  where
distinct parameters can be independently studied in order to interpret
the  experimental results.  Prominent  among them  we find  the parton
distribution  functions   which  correspond  to  the  momentum
distributions of partons inside a proton. The extraction of the parton
distribution function from  experimental data and theoretical approach
is  a complicated  task~\cite{Tung:2001cv,cteq0}.  The  uncertainties
grow  with diminishing  $x_B$ and  are generally  very much
larger for gluons  than for quarks.  At the   LHC (Large Hadron
Collider) energies,  the  gluons  will play an  important  role  to
understand  the new  physics.  Some QCD  analysis of  parton
distribution has been done~\cite{cteq0}. At this energies,
 one will  be able to explore regions of  small $x_B$ as shown
in Table~\ref{tabla1} for minimum bias events.

\begin{table}
\caption{\label{tabla1}  The $x_B$  values for  low $p_{t}$
  production at different energies.  The evaluation was done at
  central rapidity, assuming $p_{t}=500$ MeV/c. }
\begin{ruledtabular}
%\begin{tabular}{lcr}
\begin{tabular}{lcr}
$\sqrt{s}$ &$x_{B}$&Collider\\
\hline
$200$ GeV & $\sim 5\times10^{-3}$ & RHIC \\
900 GeV & $\sim 10^{-3}$ & TEVATRON/LHC \\
10  TeV & $\sim 10^{-4}$ & LHC\\
\end{tabular}
\end{ruledtabular}
\end{table}

\section{\label{Results}Results on event shape variables and PDF of gluons }
In  order to  describe the  recoil  and sphericity  variables and  their
behavior as function of the  parton distribution function of the gluon
used,   we  generated   a   sample  of   PYTHIA   6.214  minimum   bias
events~\cite{pythia}.   The generation  incorporates some  parameters
proposed as  {\it ATLAS  tune},  based on  reference~\cite{atlas}. This
tuning parameters were fixed to describe the experimental minimum bias
data of  UA5 and CDF  experiments.  The pseudorapidity  and transverse
momenta distribution  were the first variables fitted  in this tuning. Setting from these fits
after  that,  prediction  of   the  properties  of  minimum  bias  and
underlying event at LHC energy were scaled.

\begin{figure}
\resizebox{0.4\textwidth}{0.26\textheight}{\includegraphics{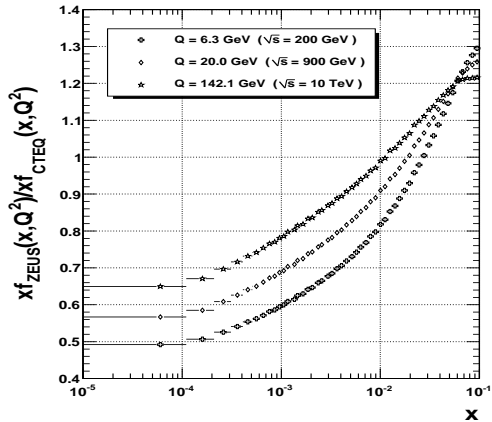}}
\caption{\label{xg} Ratio of the  gluon contributions for the ZEUS and
  CTEQ5L  parton  distribution functions,  the  ratio  is shown  for
  different Q values corresponding to  the mean Q values obtained from
  the simulations at 0.2, 0.9 and 10 TeV, respectively }
\end{figure}

We have  used this  version of PYTHIA  to implant two different parton distribution
functions:  ZEUS2005~\cite{zeus}   and  CTEQ5L~\cite{cteq}.  The
latter  PDF   was  extracted  from   a  global  analysis   to  several
experimental  data  while  the former is extracted  only  from  HERA
data. Plotting the ratio of the two gluon distribution as we have done
in Fig.~\ref{xg}, we see that at different energies and $Q$ values they
differ  substantially, the lowest  values of  $x$ showing  the largest
difference.
 
\begin{figure} %Fig. 2
%\resizebox{0.5\textwidth}{!}{\includegraphics{figures/thrustvsmeanCHnew.eps}}
\resizebox{0.4\textwidth}{0.26\textheight}{\includegraphics{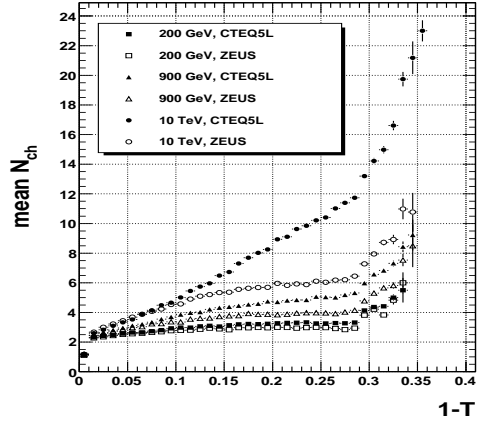}}
\resizebox{0.4\textwidth}{0.26\textheight}{\includegraphics{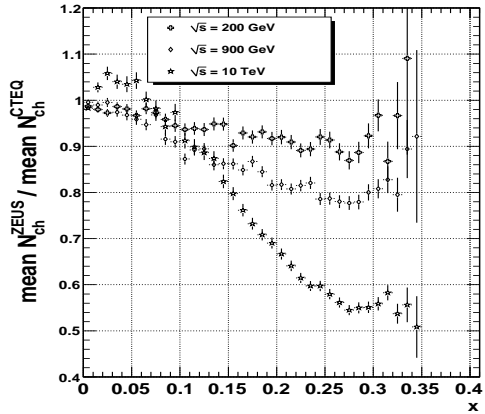}}
\caption{\label{mean-M} top: Mean  multiplicity in the pseudo rapidity
  range $\pm 0.9$ obtained  at  the three  energies  and the  two PDFs  in
  function  of  the  sphericity   variable.   Bottom: ratio  of  the  mean
  multiplicities for the  two PDF's used for the  three energies, again
  in function of the sphericity variable.}
\end{figure}

In Fig.~\ref{mean-M}  we show  in the top  panel the  mean multiplicity
variation with the sphericity variable obtained at central pseudorapidity. We
observe  an interesting dependence  on the  PDF used: the mean multiplicity at high sphericities is
much higher for  the CTEQ5L PDF than for the  ZEUS one. The phenomenon
is the most accentuated in the case of the 10 TeV collisions where the
$x_B$  value reached is  also the  smallest. The bottom part of Fig.~\ref{mean-M}
illustrates this  fact, it shows the  behavior of the ratio  of the mean
multiplicities for the three cases as a function of sphericity.

The studies  displayed here are within the  reach of all the
LHC experiments, though we have done  the simulation for the case of a
relatively  small  acceptance detector  which  has  the capability  to
detect low  charged particle momenta. We consider  only primary tracks
within $|\eta|\leq0.9$ and $p_{t}\geq0.5$ GeV$/$c defined
according to  reference~\cite{janfiete}. In this context they  are the particles
produced  in   the  collision,   including  products  of   strong  and
electro-magnetic decays, but excluding feed-down products from strange
weak decays and particles produced in secondary interactions.

\begin{figure}
\resizebox{0.4\textwidth}{0.26\textheight}{\includegraphics{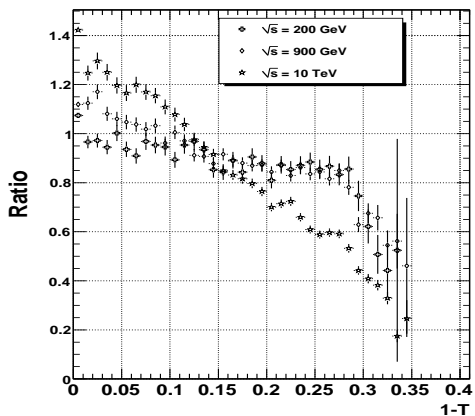}}
\caption{\label{thrust} Ratio of sphericity distributions at different energies}
\end{figure}

In   Fig.~\ref{thrust}  we   show  the   ratio  of   the  sphericity
distributions obtained  for the three  energies. One observes  that in
all cases the difference between the two PDF's increases with growing
spherificity  but   much  more   so  in  the   case  of  the   10  TeV
simulations. This is due to the  smaller $x_B$ in that case, where the
differences between the two PDF's are the largest. The strong slope in
case of 10 TeV data demonstrate the sensitivity of the approach.

\begin{figure}
\resizebox{0.4\textwidth}{0.26\textheight}{\includegraphics{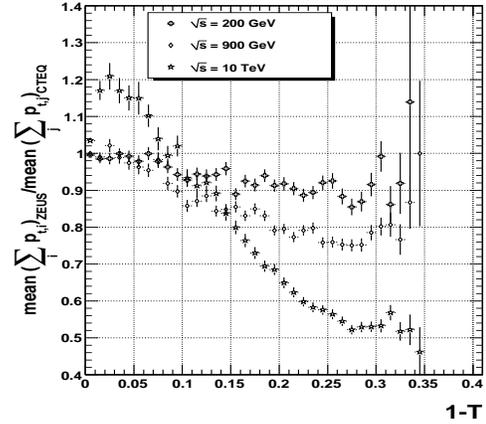}}
\caption{\label{meanpt} 
Ratio of mean total transverse momenta for the two PDFs  versus sphericity calculated for 3 energies}
\end{figure}

\begin{figure}
\resizebox{0.4\textwidth}{0.26\textheight}{\includegraphics{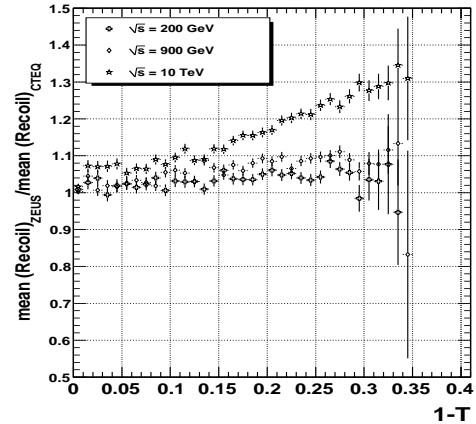}}
\caption{\label{ratiorecoil} Ratio  of  the mean
  recoil values for two PDF's versus sphericity calculated for  3 energies.}
\end{figure}

Next, the recoil variable is computed according to Eq.~\ref{ecrecoil}, the results are
shown  in Fig.~\ref{ratiorecoil} where  a characteristic  behavior of
the ratio  of the mean R  values is observed for the  3 energies. Namely,
The ZEUS  PDF has a larger  momentum inbalance than  CTEQ5L. We explain
the feature by the fact that, due  to the smaller rise of the gluon PDF
in ZEUS one is confronted with an emission of particles resulting from collisions of 
partons of  different momenta,  while in the  case of the  more peaky
gluon distribution of  CTEQ5L the chance of getting  a collision of two
partons of similar momentum is  more probable, hence resulting  in
a larger probability  that the resulting jets be  contained within the
acceptance.

\section{\label{conclusion}Conclusion}

The bulk sphericity distributions of several parameters exhibit marked
dependence on the gluon PDF used  for the computation. As the PDFs for
low  $x_B$ have  a  larger  uncertainty we  believe  that the  present
results point  to a  possibility to further  refine gluon PDFs  at low
$x_B$ using  the approach  presented.  In the  course of this  work we
have also tried to check the dependence of the parameters on different
fragmentation functions used. The results however do not demonstrate a
similar dependence  as for  the PDFs. As  one can expect,  varying the
fragmentation functions one affects slightly the absolute value of the
parameter dependence on sphericity but not the shape.\\

\noindent
In summary, the sphericity  distributions of several parameters are found to be highly
sensitive on the details of  the gluon PDF.  This opens an interesting
possibility to  compare the existing gluon PDFs  with the experimental
results using  a small sample of  minimum bias data  in pp collisions.
The Pb Pb collisions, we are currently studying~\cite {Cuautle} in the
same framework  will benefit  very much from  the present  analysis to
determine the expected gluon saturation at low $x_B$.

\begin{acknowledgments}
Support for this work has been received in part by
DGAPA-UNAM under PAPIIT grants IN115808 and IN116508 as well as by
the HELEN program.

\end{acknowledgments}

\end{document}